\documentclass[twocolumn,showpacs,preprintnumbers,amsmath,amssymb,floatfix,superscriptaddress]{revtex4}

\usepackage{epsfig}
\usepackage{graphicx} 
\usepackage{dcolumn}  
\usepackage{bm}       
\usepackage{color}
\usepackage{epstopdf}

\newcommand{\be}{\begin{equation}}
\newcommand{\ee}{\end{equation}}

\begin{document}

\title{Natural mode entanglement  as a resource for quantum communication}
\author{Libby Heaney}\email{l.heaney1@physics.ox.ac.uk}
\affiliation{Department of Physics, University of Oxford, Clarendon Laboratory, Oxford, OX1 3PU, UK}
\affiliation{Centre for Quantum Technologies, National University of
  Singapore, Singapore}
  
\author{Vlatko Vedral}
\affiliation{Department of Physics, University of Oxford, Clarendon Laboratory, Oxford, OX1 3PU, UK}
\affiliation{Centre for Quantum Technologies, National University of
  Singapore, Singapore}

\begin{abstract}
Natural particle-number entanglement resides between spatial modes in coherent ultra-cold atomic gases.  However,  operations on the modes are restricted by a superselection rule that forbids coherent superpositions of different particle numbers.  This seemingly prevents mode entanglement being used as a resource for quantum communication.   In this paper, we demonstrate that mode entanglement of a single massive particle can be used for dense coding and quantum teleportation despite the superselection rule.  In particular, we provide schemes where the dense coding linear photonic channel capacity is reached without a shared reservoir and where the full quantum channel capacity is achieved if both parties share a coherent particle reservoir.  

\end{abstract}

\maketitle

{\it Introduction} - Entanglement naturally exists within spatially coherent ultra-cold atomic gases \cite{Simon:02,Libbythesis:08}.  The indistinguishability of the atoms in such systems means, however, that no degrees of freedom can be assigned to any individual particle, which renders the concept of particle entanglement meaningless.  The natural entanglement of a BEC instead lies between second quantised field modes that are occupied by particles \cite{Simon:02, Zanardi:01}. For instance, the second quantised state of a single particle in the symmetric ground state, $\phi_0(x)$, of a confining potential written in terms of two spatial modes, $A$ and $B$, is $|\psi\rangle=\hat{a}^{\dag}_0|0\rangle=\frac{1}{\sqrt{2}}(\hat{\psi}_A^{\dag}+\hat{\psi}_B^{\dag})|0\rangle=\frac{1}{\sqrt{2}}(|1\rangle_A|0\rangle_B+|0\rangle_A|1\rangle_B)$, where $\hat{a}_0^{\dag}$ is the creation operator for  the single-particle state $\phi_0(x)$ and $\hat{\psi}^{\dag}_X$ is the field operator for spatial mode $X=A,\,B$, i.e. $\hat{\psi}^{\dag}_X|0\rangle=|1\rangle_X$. The emergence of entanglement between spatial modes is related to the BEC phase transition \cite{Anders:06} and stems from the long-range coherence properties of the gas \cite{Goold:09}. 

It is currently debated whether mode entanglement of massive particles is a useful resource in the same way as particle entanglement \cite{Wiseman:03}, i.e. EPR or spin entanglement, or whether it is just a mathematical feature of the quantum state and cannot be used at all for quantum communication and quantum information processing protocols.  
The main objection to performing such protocols using mode entanglement  of massive particles is a superselection rule that forbids the coherent superposition of different numbers of particles \cite{Wick:52}.  The superselection rule restricts quantum states to the subspace of fixed particle number and prevents spatial modes from being rotated away from the particle number basis.  However, recent theoretical results indicate that this superselection rule may be locally overcome by exchanging particles with a reservoir BEC \cite{Bartlett:07} and that mode entanglement of a single massive particle can, in principle, violate a Bell inequality \cite{Heaney:08}.  

In this paper, we illustrate that it is possible to perform quantum dense coding  \cite{Bennett:92} and teleportation \cite{Teleport} with the mode entanglement of a single massive particle.  Simple quantum information processing protocols such as these not only test specific tools needed for large scale processing, but actual experiments also serve as a bench mark for comparisons of quantum information processing with different physical realisations.  We show that the linear photonics channel capacity for dense coding \cite{Mattle:96} is, in principle, achievable without a BEC reservoir and that the full quantum channel capacity can be reached by coupling to a BEC reservoir.  Conversely, we note that hyper-entanglement, i.e. entanglement in more than one degree of freedom, will not allow to surpass the classical channel capacity limit.   
At present there has been no test for the existence of mode entanglement of massive particles,  and the predicted ability to overcome the said superselection rule remains unconfirmed.   The single mode rotations and two mode (dis)entangling gates required for this scheme can be implemented using todays technology, hence paving the way to establish, not only the existence of mode entanglement, but also whether this natural entanglement could be used for quantum information processing.  

{\it Dense coding} -  Quantum dense coding \cite{Bennett:92} allows for two classical bits of information to be transmitted via one qubit.  
A maximally entangled Bell state is shared between two parties, $A$ for Alice and $B$ for Bob, and the protocol is performed by Alice acting solely on her qubit with one of the four possible unitary operations $\{ \hat{I},\hat{Z},\hat{X},\hat{Z}\hat{X}\}$, where $\hat{X}$ and $\hat{Z}$ are the usual Pauli operators.  This encodes four classical bit patterns in the four orthogonal Bell states.  To communicate the information, Alice sends her qubit to Bob, who then performs Bell state analysis \cite{Nielsen:00}  to determine which two classical bits were initially sent.  

Dense coding was first performed in the realm of optics \cite{Mattle:96}.      
However, reliably resolving all four Bell states using linear optics alone is impossible \cite{Vaidman:99}; strong non-linear interactions are needed.   The maximum linear photonic channel capacity is $\mathcal{C}_{p}=\log_2 \,3\approx 1.585$ bits, downgraded from the quantum maximum of $\mathcal{C}_q=2$, as only two of the four Bell states can be  discriminated (the remaining two are grouped together). 
On the other hand, one can exceed the linear photonics threshold by using hyper-entangled photons \cite{Barreiro:08,Kwiat:97}. 
Unlike with photons, it is much simpler to generate an interaction between atoms, thus discriminating all four Bell states.  
A dense coding experiment with complete Bell state analysis has been performed using the continuous variable entanglement of ions \cite{Schaetz:04}.  

We consider a system of two coupled tightly confined potentials.  The tight confinement in all three directions ensures that the onsite interaction energy, $U$, is much larger than the other energy scales, so that only zero or one particle(s) can occupy each potential at any instance.  This is the case for atomic quantum dots  \cite{Recati:05}.  Each potential forms one of the spatial modes, $A$ or $B$.  There is no potential bias between the modes unless otherwise indicated.  With these conditions the full Hamiltonian, which can be derived in a similar manner to the Bose-Hubbard model for optical lattices \cite{Jaksch:04}, is simply, $\hat H = -\frac{J}{2} (\hat\psi_A^{\dag}\hat\psi_B+\hat\psi_B^{\dag}\hat\psi_A)$, where $J$ is the tunneling matrix element between the two modes.  Initially, a single particle occupies the ground state, $|\psi^+\rangle_{AB}=\frac{1}{\sqrt{2}}(|10\rangle+|01\rangle)$, of the potential, where $|01\rangle=|0\rangle_A\otimes|1\rangle_B$ denotes zero particles in mode $A$ and one particle in mode $B$.  

Consider now performing dense coding with the state, $|\psi^+\rangle$.  Since $|\psi^+\rangle$ belongs to a subspace of fixed particle number, it is impossible to rotate from it to the two Bell states, $|\phi^{\pm}\rangle=\frac{1}{\sqrt{2}}(|00\rangle\pm|11\rangle)$, without possessing a particle reservoir.  
Alice can change the phase of her mode $|\psi^+\rangle\rightarrow|\psi^{(\varphi)}\rangle=\frac{1}{\sqrt{2}}(|01\rangle+e^{i\varphi}|10\rangle)$, by adding a potential bias, $\hat{H}_A=V\hat{n}_A$ i.e. increasing the trapping potential of her mode for a certain time.  For instance, a $\hat{Z}$ rotation is performed by raising the potential by an amount $V$ relative to Bobs for the time $t=\pi/V$ (where $\hbar=1$) \cite{PhaseImprint}.  An {\it isolated} mode entangled state thus provides no quantum advantage to dense coding and gives a classical channel capacity, $C$, of unity.   

{\it Linear photonic channel capacity} - In order to overcome the superselection rules, a reservoir containing the same type of particles as the initial state is required.  Interestingly, if Alice is in possession of just a single extra particle (in addition to her mode), the linear photonic theshold for dense coding can be reached.  The $\hat{Z}$ operation can be performed as before.  To generate the third distinguishable state Alice measures her spatial mode in the particle number basis and performs a $\hat{X}=|0\rangle\langle1|+|1\rangle\langle0|$ operation on it using the extra particle.  This results in either the state, $\hat{X}_A|10\rangle=|00\rangle$ or $\hat{X}_A|01\rangle=|11\rangle$.  Alice does not need to couple to a BEC reservoir here, since one does not need to remove the knowledge about the origin of the extra particle, as one does in the full dense coding protocol below.  Also, note that the states, $|00\rangle$ and $|11\rangle$ cannot be used to encode a distinct message each, since they occur probabilistically, instead they are grouped together to encode the third message. 

After Alice completes her operation, spatial mode $A$ is sent to Bob, who then sends each mode into one of the two inputs of a 50:50 beamsplitter.  The beamsplitter transforms the modes as $\hat{\psi}^{\dag}_A=\frac{1}{\sqrt{2}}(\hat{\psi}^{\dag}_C+\hat{\psi}^{\dag}_D)$ and $\hat{\psi}^{\dag}_B=\frac{1}{\sqrt{2}}(\hat{\psi}_C^{\dag}-\hat{\psi}_D^{\dag})$.  Beamsplitters have become a standard element in atom-optics and can be realised, for instance, by controlling the exchange of particles between two potential wells for a given amount of time \cite{Sengstock:04}.  
The output states corresponding to the four possible input states are $
|\psi^+\rangle_{AB}\rightarrow|10\rangle_{CD},\,
|\psi^-\rangle_{AB}\rightarrow|01\rangle_{CD},\,
|00\rangle_{AB}\rightarrow|00\rangle_{CD},$ and $
|11\rangle_{AB}\rightarrow\frac{(|20\rangle-|02\rangle)_{CD}}{\sqrt{2}}$.
There are five distinct outcomes that indicate which of the three messages was originally sent by Alice.  The channel capacity is $\mathcal{C}_p=\log_2 3$. 
The linear photonic threshold for dense coding can be reached using mode entanglement of massive particles without a particle reservoir, even though rotations to two Bell states  are forbidden by a superselection rule. 
This result is particularly surprising given the fact that Wiseman and Vacarro \cite{Wiseman:03} previously pointed out that {\it two }copies of a mode entangled state are necessary for it to serve as a quantum resource. 

\begin{figure}[t] 
   \centering
   \includegraphics[width=3.5in]{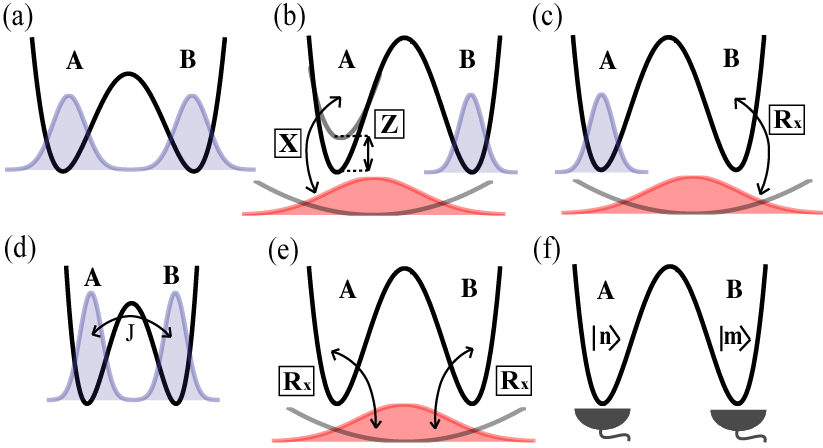} 
   \caption{ A double well potential forms the two spatial modes, $A$ and $B$. A BEC reservoir (depicted in red) shared between the two parties, is used to rotate the spatial modes  away from the particle number basis.  Steps in the dense coding protocol. {\it Initialisation}: (a)  the state is prepared in the symmetric ground state of the double well potential. {\it Encoding:} (b)  the $\hat Z$ operation is performed by adjusting the potential of mode $A$; the $\hat X$ operation by exchanging particles with a reservoir BEC. {\it Decoding:} The four steps of the Bell state analysis are (c) $\hat R_x(\pi/2)$ rotation on mode $B$ by coupling to the {\it shared} reservoir; (d) system is driven to strongly interacting limit and  the wells are coupled so that the particle(s) exchange position; (e) $\hat R_x(\pi/2)$ rotations on both modes; (f) the number of particles, $n, \,m=0\,1$, in each mode are read out. }
   \label{fig:steps}
\end{figure}

{\it Maximum quantum channel capacity} - The full dense coding protocol requires two steps, encoding, i.e. the ability to rotate to all {\it four} Bell states and decoding, i.e. {\it complete} Bell state analysis, see Fig. \ref{fig:steps}.  We will show that these steps can be implemented using mode entanglement of massive particles.  In order for the protocol to work, Alice and Bob must share a common reference frame with which they can exchange particles.  A BEC described by a mixture of coherent states, $\hat\rho_{\textrm{BEC}}=\int_0^{2\pi}\frac{d\theta}{2\pi}||\alpha|e^{i\theta}\rangle\langle|\alpha|e^{i\theta}|$, fulfills this role \cite{Dowling:06}, where $|\alpha|^2=\bar{n}$ is the average number of particles in the condensate and $\theta$ is the condensate phase.  For simplicity, we consider a non-interacting BEC.  Interactions would mean that a greater fraction of particles are out of the condensate, which would increase the probability of exchanging a particle with the system that was not phase locked between Alice and Bob.  

Recall that the initial state of the modes is $|\psi^+\rangle$ and that a $\hat Z_{A}$ rotation can be applied by creating a potential bias relative to mode $B$ for a certain time.
An $\hat{X}$ rotation requires an exchange of particles with a reservoir.  The Hamiltonian that describes this exchange is
$\hat{H}_{int}=-\frac{\Omega}{2}(\hat \psi_A^{\dag}\hat{b}_{res}+\hat{b}_{res}^{\dag}\hat{\psi}_A),$
where $\hat{b}_{res}$ is the reservoir annihilation operator.  The transition is driven by a Raman laser set-up, which couples the two different trapping states, $a$ and $b$ of the system and reservoir atoms respectively \cite{Jaksch:04}.  The parameter, $\Omega=\int dx \phi_0(x)\Psi_0(x)\tilde{\Omega}$ is the effective Rabi frequency, where $\Psi_0(x)$ is the wavefunction of the condensate, $\phi_0(x)$ is the wavefunction of the atom in mode $A$ and $\tilde{\Omega}$ is the usual Rabi frequency. The unitary evolution governed by $\hat{H}_{int}$ transforms the occupation number of the modes as follows
\begin{eqnarray}
\label{Eq:modeBEC}
|0\rangle&\rightarrow&\cos(\frac{\Omega\sqrt{\bar{n}}t}{2})|0\rangle-i\sin(\frac{\Omega\sqrt{\bar{n}}t}{2})e^{i\theta}|1\rangle\nonumber\\
|1\rangle&\rightarrow&\cos(\frac{\Omega\sqrt{\bar{n}}t}{2})|1\rangle-i\sin(\frac{\Omega\sqrt{\bar{n}}t}{2})e^{-i\theta}|0\rangle,
\end{eqnarray}
where we have traced out the reservoir (in which we also assumed a high mean number of particles, $\bar{n}>>1$).  
The state, $|\psi^+\rangle$, is rotated away from the subspace of fixed particle number and to the Bell state,  $|\phi^+\rangle_{AB}=\frac{1}{\sqrt{2}}(|00\rangle+e^{i2\theta}|11\rangle)$, by coupling to the BEC reservoir for $t=\pi/(\Omega\sqrt{\bar{n}})$.

The phase of the state, $|\phi^+\rangle$, is correlated to the BEC phase after the exchange of particles.  If Bob has no knowledge of this phase, to him the state, $|\phi^+\rangle$, would appear mixed, $\int_0^{2\pi}\frac{d\theta}{2\pi}|\phi^+\rangle\langle\phi^+|$.  However, since Bob exchanges particles with the same BEC during the Bell state analysis, the phases cancel and the full discrimination of the four Bell states is possible.  Note also that Alice and Bob must use a phase locked laser to drive the exchange of particles with the BEC.   

The Bell state analysis requires four steps comprising two different operations, namely a $\hat{R}_x(\pi/2)$ single mode rotation and a two mode disentangling c-phase type gate.   We now describe how these operations fit together to allow the four Bell states to be distinguished, see Fig. \ref{fig:steps}.

(i) Initially the potential barrier between the two modes is sufficiently high so that tunneling is suppressed.  Mode $B$ is then rotated away from the particle number basis by coupling to the BEC reservoir via the Raman laser set-up, see Eq. \ref{Eq:modeBEC}, for time $t=\pi/(2\Omega\sqrt{\bar{n}})$.  This results in the transformations, $|0\rangle_B\rightarrow\frac{1}{\sqrt{2}}(|0\rangle-ie^{i\theta}|1\rangle)_B$ and $|1\rangle_B\rightarrow\frac{1}{\sqrt{2}}(|1\rangle-ie^{-i\theta}|0\rangle)_B$.  For example, the state $|\phi^+\rangle_{AB}$ would become $|\phi^+\rangle_{AB}\rightarrow\frac{1}{2}\big{(}|0\rangle(|0\rangle-ie^{i\theta}|1\rangle)+e^{i2\theta}|1\rangle(|1\rangle-ie^{-i\theta}|0\rangle)\big{)}$.

(ii) The crucial two mode c-phase disentangling gate is implemented by driving the system into the limit of hardcore bosons, i.e. $U\rightarrow\infty$, by tightening the confining potentials further. This results in an effective zero-dimensional system 
where the bosons behave like spin-polarised fermions. By associating the two particle number states, $|0\rangle$ and $|1\rangle$, of each mode with the up/down spin half degree of freedom, the bosonic Hamiltonian, $\hat H = -\frac{J}{2}(\hat\psi_A^{\dag}\hat\psi_B+\hat{\psi}^{\dag}_B\hat{\psi}_A)$, becomes equivalent to the quantum XX spin model.  This in turn can be mapped via the Jordan-Wigner transformation to $\hat H=-J(\hat c_A^{\dag}\hat c_B+\hat c_B^{\dag}\hat c_A)$, where $\hat{c}_X^{\dag}$ and $\hat c_X$ are Fermionic creation and annihilation operators for mode, $X=A,\,B$ that anti-commute, $\{\hat c_X,\hat c_Y\}_+=\delta_{XY}$.  Once the system is in this regime, the barrier between the modes, $A$ and $B$, should be lowered for time, $t=\pi/(2J)$, so that the particles exchange position.  The two Bell states, $|\psi^{\pm}\rangle$, are invariant,  while the $|11\rangle$ term in the $|\phi^{\pm}\rangle$ states pick up a minus sign $|11\rangle\rightarrow-|11\rangle$, due to the anti-commutation relations \cite{Clark:05}.  

(iii) In order to rotate the modes to the particle number basis for the read-out, each should be coupled to the BEC reservoir for time $t= \pi/(2\Omega\sqrt{\bar{n}})$, as in step (i).  This final rotation ensures that phases picked up from the BEC are no longer present in the final state.  For example,  after the c-phase gate, $\frac{1}{2}\big{(}|0\rangle_A(|0\rangle-ie^{i\theta}|1\rangle)_B-e^{i2\theta}|1\rangle_A(|1\rangle+ie^{-i\theta}|0\rangle)_B\big{)}=\frac{1}{\sqrt{2}}(|0\rangle-ie^{i\theta}|1\rangle)_A\frac{1}{\sqrt{2}}(|0\rangle-ie^{i\theta}|1\rangle)_B$, becomes $|11\rangle_{AB}$.   The other three outcomes, $|01\rangle_{AB},\,|10\rangle_{AB},\,|00\rangle_{AB}$, also have one-to-one correspondence with the remaining three Bell states. 

(iv) Finally, each spatial mode is measured in the particle number basis.  The particle number detector should be able to distinguish between zero or one particles \cite{Schlosser:01}.

Coupling the spatial modes to a reservoir of indeterminate particle number, such as a BEC, that is phase locked between the modes, allows for total encoding and decoding.  The full quantum channel capacity can, in principle, be attained.

{\it Hyper-entangled modes for dense coding} - One may also wonder whether hyper-entanglement \cite{Kwiat:97} could be used for complete dense coding without a reservoir BEC.  Hyper-entangled particles have states that are entangled in more than one degree of freedom.  For instance, photon pairs have been entangled simultaneously in polarisation, space and time-energy degrees of freedom \cite{Barreiro:05}.  

Consider now that the single-particle in the double well set-up also possesses an internal degree of freedom such as spin. The particle could be prepared in a superposition of, for instance, spin up and spin down, $|S^+\rangle=\frac{1}{\sqrt{2}}(|\uparrow\rangle+|\downarrow\rangle)$, which when  written in terms of {\it spin modes}, results in the mode entangled state, $|S^+\rangle_{\uparrow\downarrow}=\frac{1}{\sqrt{2}}(|10\rangle_{\uparrow\downarrow}+|01\rangle_{\uparrow\downarrow})$, where $|10\rangle_{\uparrow\downarrow}=|1\rangle_{\uparrow}\otimes|0\rangle_{\downarrow}$ denotes one particle in the spin up mode and zero particles in the spin down mode.  The joint state of the spatial and spin modes is $|\Psi_1\rangle=|\psi^+\rangle_{AB}\otimes|S^+\rangle_{\uparrow\downarrow}$.  In total four orthogonal states are accessible without using a particle reservoir, the remaining three are $|\Psi_2\rangle=|\psi^-\rangle_{AB}\otimes|S^+\rangle_{\uparrow\downarrow}$, $|\Psi_3\rangle=|\psi^+\rangle_{AB}\otimes|S^-\rangle_{\uparrow\downarrow}$ and $|\Psi_4\rangle=|\psi^-\rangle_{AB}\otimes|S^-\rangle_{\uparrow\downarrow}$.  

Since there are four distinguishable states, one may expect the quantum channel capacity, $\mathcal{C}_q=2$, to be achievable.   However,  due to the nature of spatial mode entanglement, extra mode entanglement in an internal degree of freedom of the particle will not allow the classical channel capacity limit to be surpassed.  The system is prepared in the state, $|\Psi_1\rangle= |\psi^+\rangle_{AB}\otimes|S^+\rangle_{\uparrow\downarrow}$, which corresponds to the particle being, not only  in a superposition of positions, $A$ and $B$, but also in a superposition of its internal spin states, $\uparrow$ and $\downarrow$.  Dense coding requires two spatially separated parties, hence Alice is in possession of spatial mode $A$ and Bob is in position of spatial mode $B$.  Alice can act on her spatial mode to change the phase of the spatial portion of the state and/or to change the phase of the spin mode.  However, since the particle is coherently distributed over {\it both} modes, Alice can only rotate the spin state of the particle, $|S^+\rangle\rightarrow|S^-\rangle$, when the particle is present in her mode.   If Alice applies the $\hat{Z}$ operation on the particle spin, the spin and spatial modes become entangled, $|\Psi_1\rangle= |\psi^+\rangle_{AB}\otimes|S^+\rangle_{\uparrow\downarrow}\rightarrow \frac{1}{\sqrt{2}}(|10\rangle_{AB}|S^-\rangle_{\uparrow\downarrow}+|01\rangle_{AB}|S^+\rangle_{\uparrow\downarrow})$.     As spatial mode entanglement necessarily demands  particles to be coherently distributed across space \cite{Goold:09}, no party with access to just a portion of that space can rotate an internal degree of freedom of the particle all of the time.  Only two orthogonal states can be accessed here and the classical channel capacity limit cannot be exceeded.

{\it Teleportation} - The ability to perform complete Bell state analysis on mode entangled systems means that teleportation of a state of a spatial mode is also possible.
Mode $a$ is prepared in a initial state of, $|\phi\rangle_a=\alpha|0\rangle+\beta e^{i\theta} |1\rangle$, by coupling to the BEC as in Eq. \ref{Eq:modeBEC} for a chosen amount of time.   The phase of this state, relative to the BEC phase, can be changed by altering the potential of $a$.  Alice is in possession of modes, $a$ and $A$ and Bob is in possession of mode $B$.  To teleport the state of mode $a$, Alice performs a Bell state measurement on modes $a$ and $A$ by following steps (i) - (iv) from the decoding part of the dense coding protocol.  Each outcome in the particle number basis, $\{|00\rangle,\,|01\rangle,\,|10\rangle,\,|11\rangle\}$, corresponds uniquely to the correcting operation, $\{\hat{I},\,\hat{Z},\,\hat{Z}\hat{X},\,\hat{X}\}$, that Alice classically communicates to Bob.  Bob then performs this operation on his spatial mode using the {\it same} shared reservoir BEC to recover the original state.  The phase of the reservoir BEC is present in the final teleported state even after Bob has made his correction, which means that a third party would require a portion of the BEC to use the state any further.

{\it Conclusion} - Spatial mode entanglement of massive particles is a useful resource for quantum communication despite the superselection rule.  To reach the full quantum channel capacity for dense coding and to faithfully perform teleportation, a reservoir BEC must be shared between both parties.   Unlike with photons, full Bell state analysis can be performed utilizing the strong non-linear interactions of cold-atoms.  Even without a BEC reservoir reaching the linear photonic channel capacity for dense coding is still possible.  Conversely, hyper-entanglement of different modes does not allow to surpass the classical channel capacity.  All aspects of this scheme can be implemented using current state-of-the-art technology.

{\it Acknowledgements} - L.H. would like to thank T. Rudolph and D. Jaksch for valuable discussions.  L.H. acknowledges the financial support of EPSRC, UK. V.V. is grateful for funding
from the National Research
Foundation and the Ministry of Education
(Singapore).


\begin{thebibliography}{99}

\bibitem{Simon:02} 
	Ch. Simon, 
	Phys. Rev. A, {\bf 66} 052323 (2002).

\bibitem{Libbythesis:08} 
	L.~Heaney,
 	Ph.D thesis, University of Leeds (2008).

\bibitem{Zanardi:01}
	P. Zanardi, 
	Phys. Rev. A, {\bf65},  042101 (2001).

\bibitem{Anders:06}
	J. Anders, {\it et al.} 
	N. J. Phys., {\bf 8} 140 (2006).
		
\bibitem{Goold:09}
	J. Goold, L. Heaney, Th. Busch and V. Vedral,
	Phys. Rev. A {\bf 80} 022338 (2009).
	L. Heaney, J. Anders, D. Kaszlikowski and V. Vedral, 
	Phys. Rev. A, {\bf 76}, 053605 (2007).
	
\bibitem{Wiseman:03} 
	H.M. Wiseman and J. A. Vaccaro,
  	Phys.~Rev.~Lett.,~{\bf 91} 097902 (2003).
	
\bibitem{Wick:52} 
	G. C. Wick, A. S. Wightman, and E. P. Wigner,
	Phys. Rev., {\bf 88} 101 (1952).  
	
\bibitem{Bartlett:07}
	S.D.~Bartlett, T.~Rudolph and R.W.~Spekkens,
	Rev.~Mod.~Phys.,~{\bf 79} 555 (2007).

\bibitem{Heaney:08}
	L. Heaney and J. Anders,
	Phys. Rev. A, {\bf 80} 032104 (2009).

\bibitem{Bennett:92}
	C. H. Bennett and S. J. Wiesner, 
	Phys. Rev. Lett., {\bf 69} 2881 (1992).

\bibitem{Teleport}
	C.H. Bennett, {\it et al.},
	Phys. Rev. Lett., {\bf 70} 1895 (1993).

\bibitem{Mattle:96}
	K. Mattle, {\it et al.},
	Phys. Rev. Lett., {\bf 76} 4656 (1996).


\bibitem{Nielsen:00}
	M. A. Nielsen and I. Chuang, 
	{\it Quantum computation and quantum information}, 
	Cambridge University Press (2000).


\bibitem{Vaidman:99} 
	L. Vaidman and N. Yoran, 
	Phys. Rev. A, {\bf 59} 116 (1999).
	
\bibitem{Barreiro:08} 
	J. T. Barreiro, T-C. Wei and P. G. Kwiat, 
	Nature Physics, {\bf 4} 282 (2008).
	
\bibitem{Kwiat:97}
	P. G. Kwiat,
	J. Mod. Opt., {\bf 44} 2173 (1997).
	
\bibitem{Schaetz:04}
	T. Schaez, {\it et al.}
	Phys. Rev. Lett., {\bf 93} 040505 (2004).
	
\bibitem{Recati:05}
	A. Recati, {\it et al.}
	Phys. Rev. Lett., {\bf 94} 040404 (2005).
	
\bibitem{Dowling:06}  
	Formally, we treat the BEC as the state, $||\alpha|e^{i\theta}\rangle$, and to obtain the true quantum state of the laboratory, i.e. $\hat \rho_{\textrm{BEC}}$ + system, apply the twirling operator as defined in e.g.,	M. R. Dowling, {\it et al.},
	Phys. Rev. A, {\bf 74} 052113 (2006).

\bibitem{Jaksch:04}  	D. Jaksch and P. Zoller,
	Ann. Phys., {\bf 315}, 52 (2005).	 

\bibitem{PhaseImprint} 
	Note that controlling the phase of a BEC in this way is known as phase imprinting; see e.g.,
	J. Denschlag {\it et al.}
	Science, {\bf 287} 97 (2000). 

\bibitem{Sengstock:04} 
	K.~Bongs and K.~Sengstock, 
	Rep.~Prog.~Phys., {\bf 67}, 907 (2004).
	
\bibitem{Clark:05}
	S. Clark, {\it et al.}
	N. J. Phys., {\bf 7} 124 (2005).

\bibitem{Schlosser:01}
	N. Schlosser {\it et al.},
	Nature, {\bf 411} 1024 (2001); W. Bakr {\it et al.} arXiv:0908.0174v1.

\bibitem{Barreiro:05}
	J. T. Barreiro, {\it et al.},	Phys. Rev. Lett., {\bf 95} 260501 (2005).



	
\end{thebibliography}
\end{document}